\documentclass[aps,prb,reprint,amsmath,amssymb,superscriptaddress]{revtex4-1}

\usepackage{bm}
\usepackage{graphicx}

\renewcommand{\vec}[1]{\boldsymbol{#1}} 

\begin{document}

\title{Kondo Effect with Weyl Semimetal Fermi Arcs}

\author{Da Ma}
\affiliation{International Center for Quantum Materials, School of Physics, Peking University, Beijing 100871, China}
\author{Hua Chen}
\affiliation{Department of Physics, Zhejiang Normal University, Jinhua 321004, China}
\author{Haiwen Liu}
\affiliation{Center for Advanced Quantum Studies, Department of Physics, Beijing Normal University, Beijing 100875, China}
\author{X. C. Xie}
\affiliation{International Center for Quantum Materials, School of Physics, Peking University, Beijing 100871, China}
\affiliation{Collaborative Innovation Center of Quantum Matter, Beijing 100871, China}
\date{\today{}}

\begin{abstract}
We investigate the Kondo effect of the Fermi arcs in a time-reversal-invariant Weyl semimetal with the variational method. To show the consequence brought out by the nontrivial spin texture, we calculate the spatial spin-spin correlation functions. The correlation functions exhibit high anisotropy. The diagonal correlation functions are dominated by the antiferromagnetic correlation while the off-diagonal part has more complicated pattern. The correlation functions obey the same symmetry as the spin texture. Tuning chemical potential changes the pattern of the correlation functions and the correlation length. The correlation functions of the Weyl semimetal Fermi arcs and that from a Dirac semimetal show discrepancy.
\end{abstract}


\maketitle

\section{Introduction}
\label{intro}
Recent years, Weyl semimetals have attracted a lot of attention.\cite{weyl1929elektron,PhysRevB.83.205101,vafek2014dirac} Weyl semimetals are semimetals whose touching bands can be described by the Weyl equation around the touching points. Either breaking the time-reversal symmetry or the inversion symmetry of a Dirac semimetal, a Weyl semimetal is obtained. To date, inversion-symmetry-breaking Weyl semimetals have been found in the transition metal monopnictides,\cite{Huang:2015aa,PhysRevX.5.031013,yang2015weyl,xu2015discovery} while time-reversal-symmetry-breaking ones remain elusive.\cite{PhysRevLett.107.186806,PhysRevLett.107.127205,PhysRevB.85.165110,hosur2013recent,PhysRevLett.113.187202,PhysRevLett.117.236401} Besides hosting Weyl fermions,\cite{lv2015observation} Weyl semimetals have many interesting features such as nontrivial surface states called Fermi arcs,\cite{PhysRevB.83.205101,Huang:2015aa,PhysRevX.5.031013,xu2015discovery} chiral anomaly,\cite{nielsen1983adler,PhysRevB.87.235306,PhysRevB.88.104412,PhysRevX.5.031023,Lu2016} unusual quantum oscillations originated from Fermi arcs.\cite{potter2014quantum,moll2016transport} Other proposed phenomena include possible emergent supersymmetry,\cite{PhysRevLett.114.237001} Imbert-Federov shift,\cite{PhysRevLett.115.156602,PhysRevB.93.195165} and disorder-induced novel phase transitions.\cite{PhysRevLett.115.246603,PhysRevLett.116.066401}

When interacting with magnetic impurities, Weyl semimetals bring out new physics. The Kondo effect of the Weyl semimetal bulk states has been studied. While time-reversal-invariant Weyl semimetals as well as Dirac semimetals belong to the pseudogapped Kondo case, numerical renormalization group calculation reveals that the perturbed system shows unconventional Kondo physics.\cite{PhysRevB.92.121109} The interplay of Kondo effect and long-range scalar disorder results in non-Fermi liquid behavior.\cite{PhysRevB.92.041107} A variational study calculated the spatial spin-spin correlation functions and distinguished a Dirac semimetal from a Weyl semimetal.\cite{PhysRevB.92.195124} For multi-impurity case, the Ruderman-Kittel-Kasuya-Yosida (RKKY) interaction has also been discussed.\cite{PhysRevB.92.224435,PhysRevB.92.241103} However, those works mainly concentrate on the bulk states of the topological semimetals. The Kondo effect of the surface states in the Weyl and Dirac semimetals, Fermi arcs, has not been studied. Connecting Weyl points of opposite chirality, Fermi arcs are disjoint Fermi surfaces and have rich spin texture. As a result of inversion symmetry breaking, the shape and the spin texture of surface states in Weyl semimetals is more complicated than that of topological insulators and Dirac semimetals.\cite{PhysRevB.92.115428,PhysRevLett.115.217601,PhysRevLett.116.096801} The unique spin texture of Fermi arcs has its special impact on the Kondo effect.

Here we focus on the Kondo effect of the Fermi arcs in a time-reversal-invariant Weyl semimetal. Specifically, we study the influence of the spin texture on the Kondo effect. We solve the Anderson model with the variational method to calculate the spatial spin-spin correlation functions. We take a trial wavefunction where the impurity spin is fully compensated and find that it has lower energy than the non-interacting ground state, i.e. the impurity tends to be screened by the conducting electrons in the Fermi arcs. The spatial spin-spin correlation functions are highly anisotropic and they have the same symmetry as the Fermi arcs. Tuning chemical potential and changing the length of the Fermi arcs, the evolution of the correlation functions reveals how they are related to the details of the Fermi arcs. We also compare the spin-spin correlation functions of the Fermi arcs in the Weyl semimetal with that of the Fermi arcs in the Dirac semimetal Na$_{3}$Bi. It turns out that the relatively simpler spin texture and the shape of the Fermi arcs in the Dirac semimetal results in less anisotropy in the spatial spin-spin correlation functions with less structure, which makes it possible to distinguish the Fermi arc in a Dirac semimetal from that in a Weyl semimetal.

This article is organized as follows: In Sec.\ref{moHa}, the model Hamiltonian is presented, which describes a magnetic impurity on the surface of a Weyl semimetal. The variational method is introduced in Sec.\ref{vame} and the binding energy is calculated there. In Sec.\ref{sccl}, the spatial spin-spin correlation functions are studied based on the trial wavefunction. Comparison to Dirac semimetal Fermi arcs is made in Sec.\ref{cds}. Finally, Sec.\ref{codi} contains the conclusion and discussion part.

\section{model Hamiltonian}
\label{moHa}
To study the interaction between the Fermi arcs and the magnetic impurity, we consider the Anderson model,
\begin{equation}
H=H_{c}+H_{mix}+H_{d}.
\end{equation}
$H_{c}$ describes the Fermi arcs in the Weyl semimetal, with $c^{\dagger}_{\vec{k}}$ creates a state of momentum $(k_{x}, k_{y})$,
\begin{equation}
H_{c}=\sum (\epsilon(\vec{k})-\mu) c^{\dagger}_{\vec{k}}c_{\vec{k}}.
\end{equation}
Note that the states in the Fermi arcs are fully spin-polarized. Namely, $c^{\dagger}_{\vec{k}}$ creates a state with a certain spin polarization. $H_{mix}$ describes the interaction between the impurity and the surface states,
\begin{equation}
H_{mix}=\sum V_{\vec{k}}c^{\dagger}_{\vec{k}}d_{\vec{k}} + h.c. .
\end{equation}
We suppose the hopping conserves spin, i.e. the state created by $d^{\dagger}_{\vec{k}}$ has the same spin polarization as $c^{\dagger}_{\vec{k}}$. We will show that the spin is polarized on the $x-y$ plane in the following Fermi arcs model. As a result,
\begin{equation}
\label{eq:dk}
d^{\dagger}_{\vec{k}}=\frac{1}{\sqrt{2}}(e^{-i\frac{\phi_{\vec{k}}}{2}}d^{\dagger}_{\uparrow}+e^{i\frac{\phi_{\vec{k}}}{2}}d^{\dagger}_{\downarrow}),
\end{equation}
where $\phi_{\vec{k}}$ is the polar angle of the in-plane spin polarization with respect to the $x$-axis such that $\tan \phi_{\vec{k}} = \frac{D_{y}(\vec{k})}{D_{x}(\vec{k})}$. The last term in the Hamiltonian describes the impurity with an on-site Hubbard repulsion,
\begin{equation}
H_{d}=\sum_{\sigma}(\epsilon_{d}-\mu)d^{\dagger}_{\sigma}d_{\sigma} + Un_{\uparrow}n_{\downarrow},
\end{equation}
where we take $\epsilon_{d} = -0.3t$. For the dispersion and spin texture of the Weyl semimetal Fermi arcs, we use the result from a tight-binding model in a zinc-blende lattice.\cite{PhysRevB.87.245112} The model contains a Weyl semimetal phase which has 12 inequivalent Weyl points in the first Brillouin zone. All 12 Weyl points locate at $\epsilon(\vec{k})=0$. The Weyl semimetal breaks the inversion symmetry but preserves time-reversal symmetry. On $(001)$-plane, it is shown that the dispersion of the surface states is\cite{PhysRevB.87.245112}
\begin{equation}
\epsilon(\vec{k}) = 4t\sin{\frac{k_{x}a}{4}}\sin{\frac{k_{y}a}{4}},
\end{equation}
while $k_{x}, k_{y}$ satisfy
\begin{equation}
D(\vec{k}) > \epsilon_{0} > 0,
\end{equation}
where
\begin{equation}
D_{x}(\vec{k})=\lambda [-2\sin{\frac{k_{x}a}{2}}-\sin{\frac{a(k_{x}+k_{y})}{2}}-\sin{\frac{a(k_{x}-k_{y})}{2}}],
\end{equation}
\begin{equation}
D_{y}(\vec{k})=\lambda [2\sin{\frac{k_{y}a}{2}}+\sin{\frac{a(k_{x}+k_{y})}{2}}+\sin{\frac{a(k_{y}-k_{x})}{2}}],
\end{equation}
and
\begin{equation}
D(\vec{k})=\sqrt{D_{x}^{2}(\vec{k})+D_{y}^{2}(\vec{k})}.
\end{equation}
\begin{figure}
\includegraphics[width=8.cm]{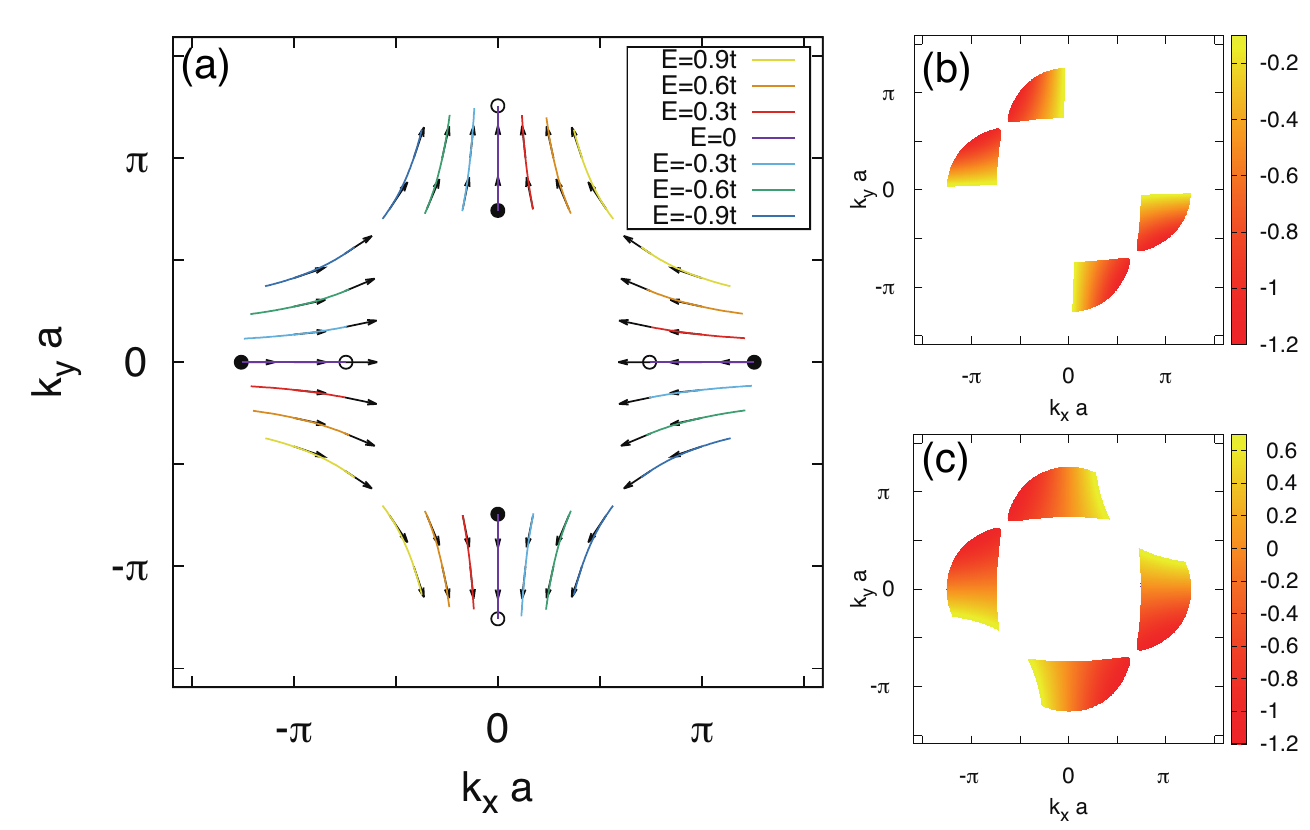}
\caption{(a) Surface states for $\vert \frac{\epsilon_{0}}{4\lambda}\vert=0.92$. The black arrows show the spin polarization. The surface states have four branches and are spin-momentum locked. Two-fold rotation symmetry is obeyed. The black and white circles show the Weyl points of opposite chirality. (b) Filled surface states for $\mu=-0.1t$. The color indicates the energy of the surface states. The corresponding spatial spin-spin correlation functions are shown in Fig.\ref{fig:165}. (c) Same as (b), but $\mu=0.7t$. Spin-spin correlation functions shown in Fig.\ref{fig:164}.}\label{fig:arcs092}
\end{figure}
In the original model Hamiltonian of the Weyl semimetal, $t$ is the nearest hopping strength, $\lambda$ is the spin-orbit coupling strength of next-nearest neighbors, and $\epsilon_{0}$ is the on-site potential which serves as the inversion symmetry breaking term.\cite{PhysRevB.87.245112} The surface states are spin-momentum locked. The spin polarization of the surface state with momentum $\vec{k}=(k_{x}, k_{y})$ is $(-D_{x}, -D_{y}, 0)$. In the model, the spin of the surface states lies in-plane. The surface states obey a two-fold rotational symmetry, see Fig.\ref{fig:arcs092}(a).
\section{Variational Method}
\label{vame}
To solve the Kondo screening problem in the Fermi arcs above, we apply the variational approach.\cite{PhysRev.147.223,PhysRevB.13.2950} The variational approach has been used as a non-perturbative way to study the Kondo screening in topological insulators, \cite{PhysRevB.81.235411} Dirac/Weyl semimetals\cite{PhysRevB.92.195124} and other systems.\cite{feng2011kondo} Following the standard procedure, first we take the ground state without the impurity as
\begin{equation}
\label{eq:ground1}
|\Psi_{0} \rangle = \prod_{\epsilon(\vec{k})<\mu}c^{\dagger}_{\vec{k}} |0\rangle,
\end{equation}
where the product runs over all occupied states below Fermi energy. Then we construct a trial wavefunction when the impurity is present. Here we consider the case where the impurity state is singly-occupied. The chemical potential $\mu$ lies between the two energy levels $\epsilon_{d}$ and $\epsilon_{d} + U$. We suppose the impurity moment is fully compensated as we use the following ansatz
\begin{equation}
|\Psi \rangle = (a_{0} + \sum_{\epsilon(\vec{k})<\mu}a_{\vec{k}}d^{\dagger}_{\vec{k}}c_{\vec{k}}) |\Psi_{0}\rangle
\end{equation}
as trial wavefunction. $d^{\dagger}_{\vec{k}}$ creates an electron in the impurity atom with the same spin-polarization as the electron annihilated by $c_{\vec{k}}$, see Eq.(\ref{eq:dk}). The energy for this trial wavefunction satisfies:
\begin{equation}
E = \frac{\sum_{\epsilon(\vec{k})<\mu}\left[ (\epsilon (\vec{k})-\mu) a_{0}^{2} + (E_{0}-\epsilon(\vec{k})+\mu)a_{\vec{k}}^{2}+ 2V_{\vec{k}}a_{0}a_{\vec{k}}\right]}{a_{0}^{2}+\sum_{\epsilon(\vec{k})<\mu}a_{\vec{k}}^{2}},
\end{equation}
where $E_{0}$ denotes the energy for the ground state in Eq.(\ref{eq:ground1}) with the impurity state singly-occupied, i.e.,
\begin{equation}
E_{0} = \epsilon_{d}-\mu + \sum_{\epsilon(\vec{k})<\mu} (\epsilon(\vec{k})-\mu).
\end{equation}
The binding energy is defined as $\Delta = E_{0}-E$. When the binding energy is positive, our trial wavefunction is preferred against $ \vert \Psi_{0}\rangle$.
Variational principle dictates $\frac{\partial E}{\partial a_{0}}=0$ and $\frac{\partial E}{\partial a_{\vec{k}}}=0$. These yield
\begin{equation}
\sum_{\epsilon(\vec{k})<\mu} a_{\vec{k}} V_{\vec{k}} = (E - \sum_{\epsilon(\vec{k})<\mu} (\epsilon(\vec{k})-\mu)) a_{0},
\end{equation}
and
\begin{equation}
\label{eq:aka0}
(E_{0} - E - \epsilon(\vec{k})+\mu) a_{\vec{k}} = -V_{\vec{k}} a_{0}.
\end{equation}
The two equations above combine to give an equation for binding energy $\Delta$:
\begin{equation}
\sum_{\epsilon(\vec{k})<\mu} \frac{V_{\vec{k}}^{2}}{\epsilon(\vec{k})-\mu -\Delta} = \epsilon_{d}-\mu - \Delta.
\end{equation}
Replacing the summation with an integral, we get a self-consistent integral equation. Numerical calculation shows that the binding energy has a positive solution for arbitrary finite coupling strength $V_{\vec k}$, which justifies the trial wavefunction. Note that for the bulk states there is a critical coupling strength when the Femi energy is at the Dirac/Weyl point. However, for the surface states there is no critical $V_{\vec k}$.
\section{Screening Cloud}
\label{sccl}
The trial wavefunction in the last section contains a lot of information about the behavior of a magnetic impurity on the surface of a Weyl semimetal. Since we are most interested in the effect originating from the nontrivial spin texture, we will focus on the screening cloud here and calculate the spatial spin-spin correlation functions between the conducting electrons in the Fermi arcs and the impurity using the trial wavefunction. At the presence of translational invariance, we can take the impurity site as the origin and define the spin-spin correlation function as follows:
\begin{equation}
J_{uv}(\vec r) = \langle S_{c}^{u}(\vec r)S_{d}^{v}(0)\rangle.
\end{equation}
Here c stands for conducting electrons in the Fermi arcs, d for the impurity electrons, $u$ and $v$ for spin indices. In this case,
\begin{equation}
\label{eq:generalJuv}
J_{uv}(\vec r) = -\frac{1}{4} \sum_{\substack{\epsilon(\vec{k_{1}})<\mu \\ \epsilon(\vec{k_{2}})<\mu}} a_{\vec{k_{1}}} a_{\vec{k_{2}}} e^{i (\vec{k_{1}}-\vec{k_{2}})\cdot \vec{r}} \gamma^{\dagger}_{\vec{k_{1}}} \sigma_{u} \gamma_{\vec{k_{2}}} \gamma^{\dagger}_{\vec{k_{2}}} \sigma_{v} \gamma_{\vec{k_{1}}},
\end{equation}
\begin{figure} 
\includegraphics[width=8cm]{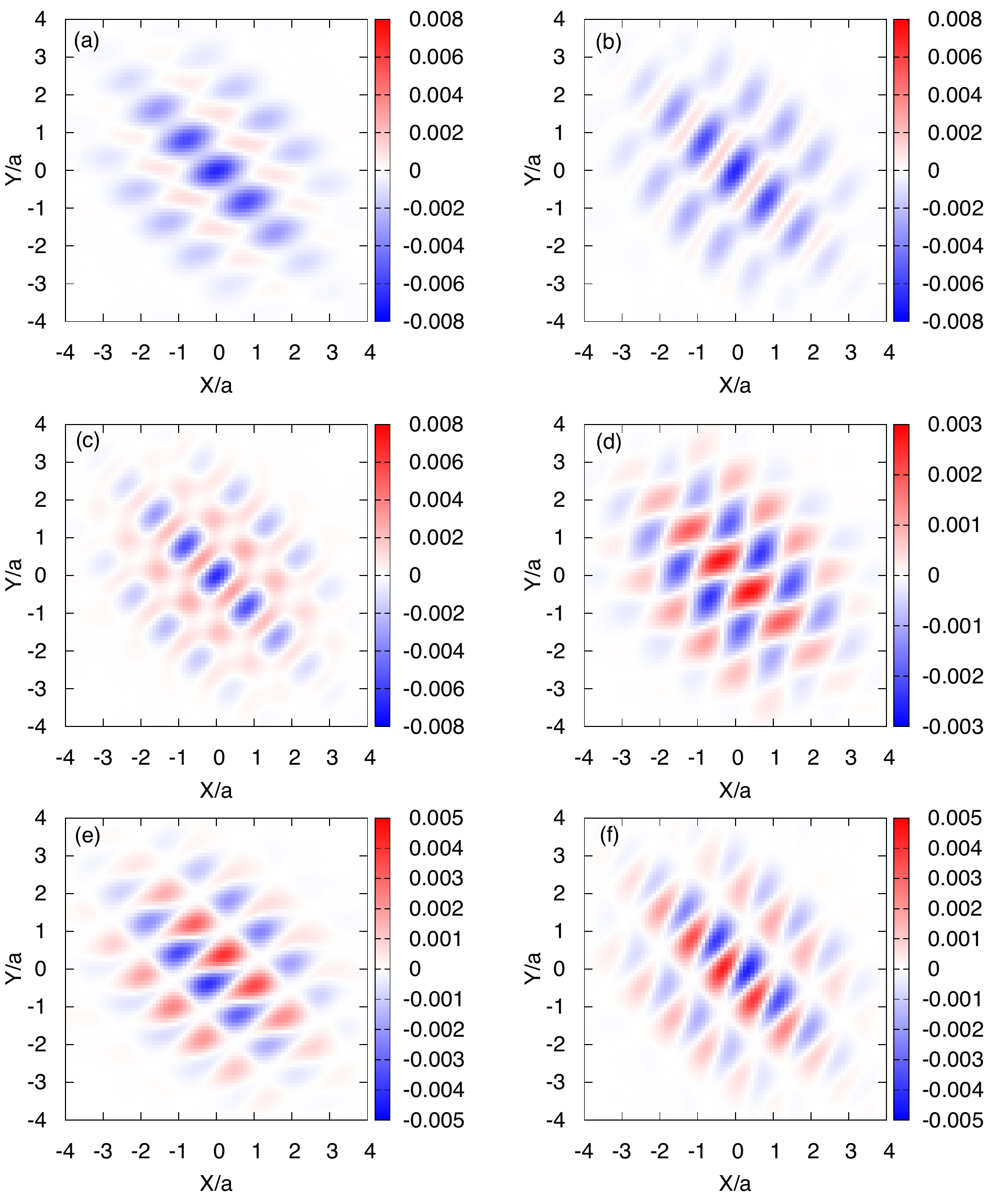}
\caption{Spin-spin correlation functions (a) $J_{xx}$ (b) $J_{yy}$ (c) $J_{zz}$ (d) $J_{xy}$ (e) $J_{xz}$ and (f) $J_{yz}$ with the chemical potential set near the Weyl points. $\vert \frac{\epsilon_{0}}{4\lambda} \vert=0.92, \mu = -0.1t, \Delta = 1.36t, V_{\vec k}=2.5t,$ and $\epsilon_{d} = -0.3t$. $X$ and $Y$ are spatial coordinates in unit of lattice constant $a$. Other off-diagonal terms are $J_{yx}(\vec r)=J_{xy}(\vec r)$, $J_{zx}(\vec r)=-J_{xz}(\vec r)$, and $J_{zy}(\vec r)=-J_{yz}(\vec r)$. The spin-spin correlation function shows anisotropy. All correlation spots are distributed from left top to right bottom, similar to the distribution of the filled surface states $\epsilon(\vec{k})<\mu$ (See Fig.\ref{fig:arcs092}(b)).}\label{fig:165}
\end{figure}
\begin{figure}
\includegraphics[width=8cm]{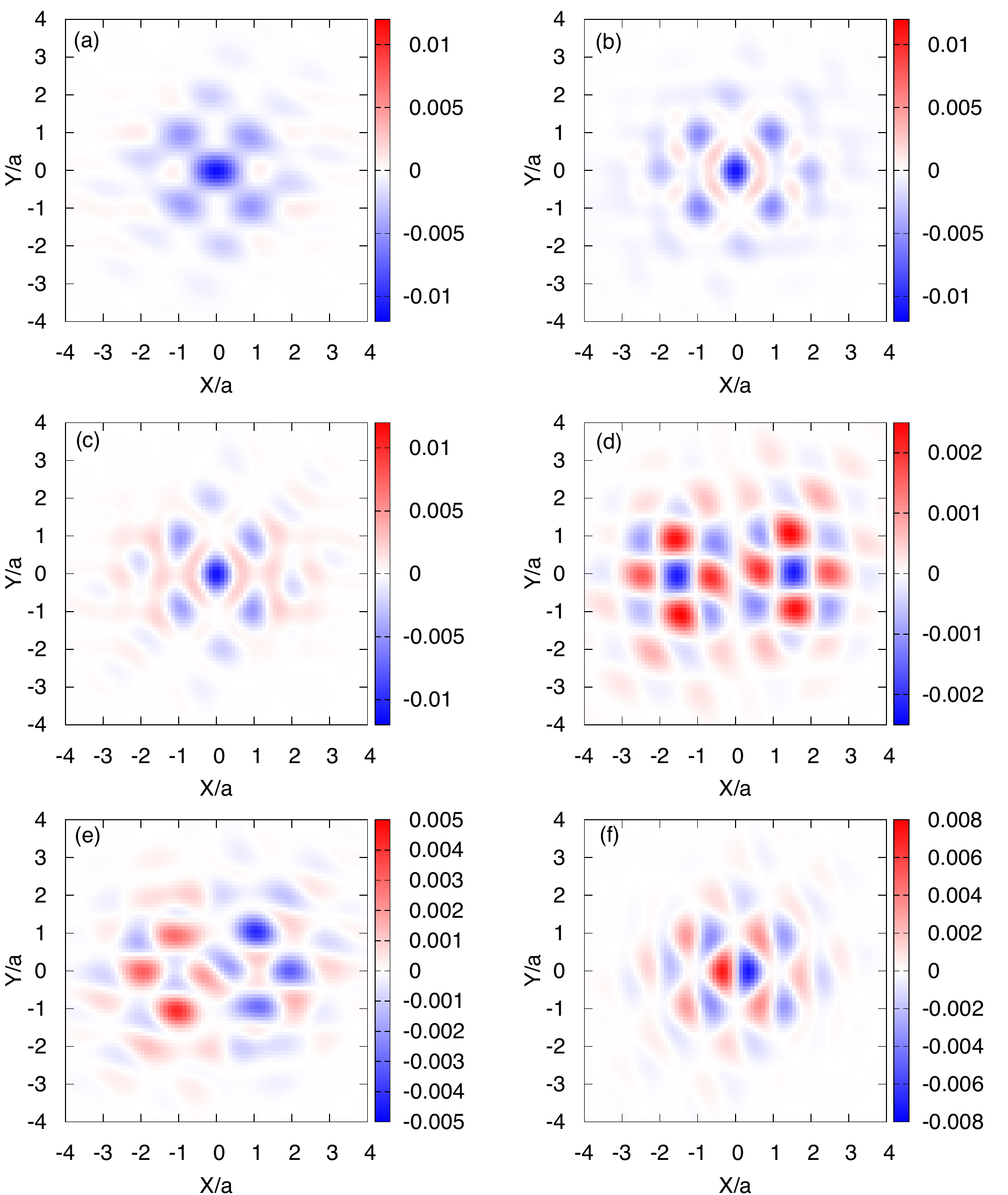}
\caption{Spin-spin correlation functions (a) $J_{xx}$ (b) $J_{yy}$ (c) $J_{zz}$ (d) $J_{xy}$ (e) $J_{xz}$ and (f) $J_{yz}$, with a higher chemical potential than Fig.\ref{fig:165} and more states filled (See Fig.\ref{fig:arcs092}(c)). $\vert \frac{\epsilon_{0}}{4\lambda} \vert=0.92, \mu = 0.7t, \Delta = 1.37t, V_{\vec k}=2.5t,$ and $\epsilon_{d} = -0.3t$. $X$ and $Y$ are spatial coordinates in unit of lattice constant $a$. Compared to the $\mu = -0.1t$, the chemical potential is raised and most $E>0$ surface states are filled. As a result, the correlation spots are no longer restricted along the line from left top to right bottom.}\label{fig:164}
\end{figure}
where $\gamma_{\vec{k}} = \frac{1}{\sqrt{2}} \left(\begin{array}{cc} e^{-i \frac{\phi_{\vec{k}}}{2}},  e^{i \frac{\phi_{\vec{k}}}{2}}\end{array}\right)^{T}$. Other off-diagonal correlation functions obey $J_{yx}(\vec r) = J_{xy}(\vec r)$, $J_{zx}(\vec r) = -J_{xz}(\vec r)$, and $J_{zy}(\vec r) = -J_{yz}(\vec r)$. Note that a minus sign appears when we interchange the spin indices $u$ and $v$ for off-diagonal terms containing $z$-spin. This can be seen from Eq.(\ref{eq:generalJuv}) as a direct consequence of the in-plane polarization. This feature provides a way to test if the spin of the Fermi arcs lies in a plane.

A set of typical spatial spin-spin correlation functions is shown in Fig.\ref{fig:165}, where the chemical potential is set slightly below the Weyl points. The correlation functions are highly anisotropic, since the six correlation functions shown are all different and none of the six graphs is of circular shape. This reflects the anisotropy in the shape and the spin texture of the Fermi arcs.

The diagonal spatial spin-spin correlation functions consist of a series of antiferromagnetic peaks, in accordance with the spin screening picture. However, there are also some small regions of ferromagnetic correlation. For the off-diagonal part, both parallel and antiparallel correlation are present, as the result of the complex spin texture for Fermi arcs.

The spin texture in our Fermi arcs model has $C_{2}$ symmetry. The symmetry is also reflected in the spatial spin-spin correlation functions. $J_{xx}$, $J_{yy}$, $J_{zz}$ and $J_{xy}$ are unchanged after the rotation, while $J_{xz}$ and $J_{yz}$ get a minus sign due to different representation. The reason is that our system respects the time-reversal symmetry and the $z$-spin axis has to be flipped after a rotation by angle $\pi$.

Raising the chemical potential changes the correlation functions in two ways. Compare Fig.\ref{fig:165} with Fig.\ref{fig:164} where the chemical potential is higher. One change is about the correlation length. As the chemical potential rises, more states in the Fermi arcs take part in the Kondo screening and the screening cloud turns less extended in space. Another change reflects the influence of the shape of the Fermi arcs.  From Eq.(\ref{eq:aka0}) we know that $a_{\vec{k}} = -\frac{V_{\vec{k}}}{\Delta - (\epsilon(\vec{k})-\mu)} a_{0}$, which indicates that the major contribution to the correlation function is made by the states near the Fermi level. In Fig.\ref{fig:165}, the correlation pattern in all the spin-spin correlation functions is distributed mainly from left top to the right bottom, as the Fermi arcs below the chemical potential sit along the same direction in $k$-space, see Fig.\ref{fig:arcs092}(b). In Fig.\ref{fig:164}, the chemical potential is raised and most of the $E>0$ sector is filled in addition to filled $E<0$ sector (See Fig.\ref{fig:arcs092}(c)). As a result, the surface states are now not restricted to the second quadrant and the fourth quadrant in $k$-space. The diagonal spin-spin correlation functions show four smaller spots at the four corners of the central spot. The off-diagonal part is also freed from that restriction.
\begin{figure}
\includegraphics[width=8.5cm]{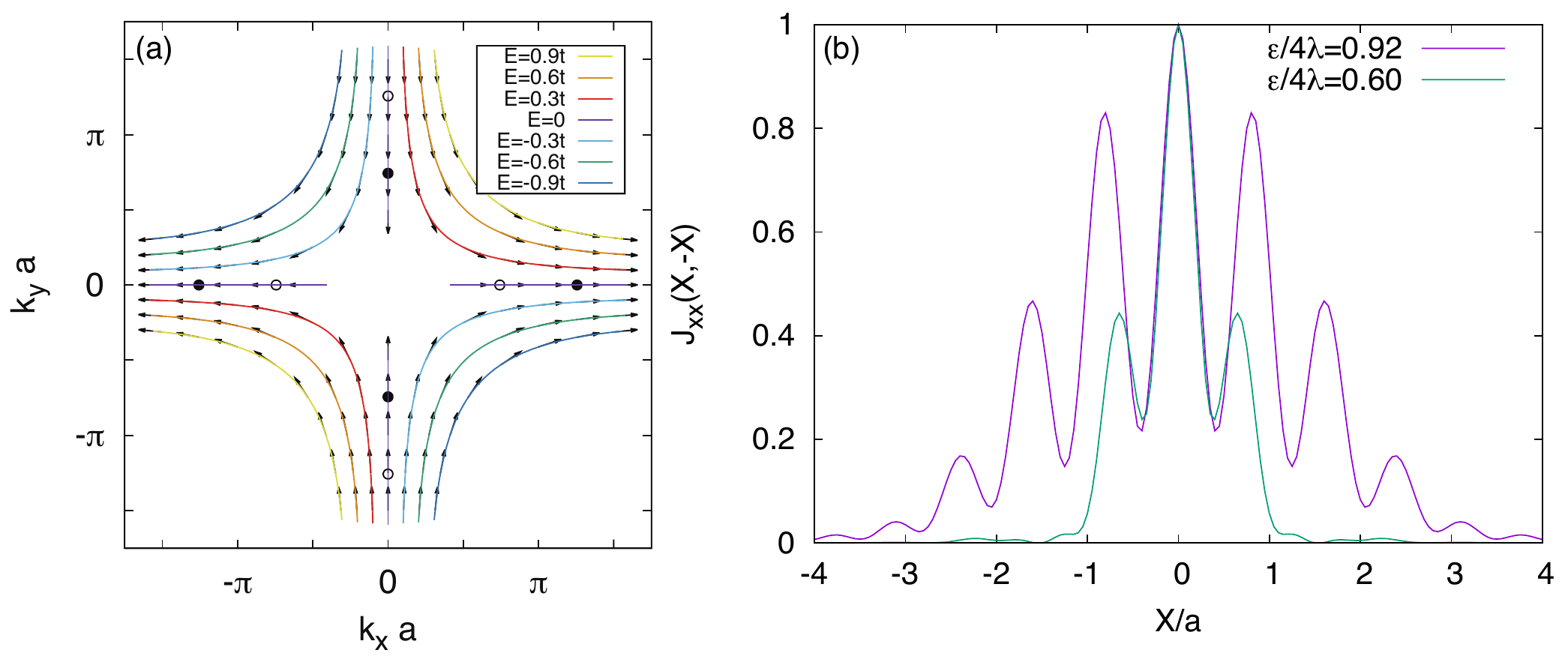}
\caption{(a) Surface states for $\vert \frac{\epsilon_{0}}{4\lambda}\vert=0.6$. The black arrows show the spin polarization. The black and white circles show the Weyl points of opposite chirality. Smaller $\vert \frac{\epsilon_{0}}{4\lambda}\vert$ brings the model deeper into the Weyl semimetal phase, and the Fermi arcs grow longer. The four pieces of surface states merge into two above (below) a certain energy for the $E>0$ ($E<0$) sector. (b) The spin-spin correlation function $J_{xx}$ along $X=-Y$ for different Fermi arcs length. We compare the arcs in Fig.\ref{fig:165} (with the same parameter there) with the arcs in (a) (Parameters: $\vert\frac{\epsilon_{0}}{4\lambda} \vert=0.6, \mu = -0.1t, \Delta = 1.1t, V_{\vec k}=2.5t$). As the Fermi arcs grow longer, the Kondo screening cloud tends to be less spatially extended. For better comparison, we have set the highest value to $1$ in this figure.}\label{fig:varlength}
\end{figure}
Tuning $\vert \frac{\epsilon_{0}}{4\lambda} \vert$, the length of the surface states is varied. When we set $\vert \frac{\epsilon_{0}}{4\lambda} \vert=0.6$, smaller than the previous value $0.92$, the Fermi arcs grow longer and two pieces of surface states connect below certain energy level, see Fig.\ref{fig:varlength}(a). For comparison, we choose a similar chemical potential and binding energy where the $E<0$ part is filled and makes the major contribution. The pattern of the correlation functions changes little, and the major difference comes in the correlation length, see Fig.\ref{fig:varlength}(b). Similar to the situation above, the longer Fermi arcs have more states participating in the Kondo screening process, which makes the correlation length shorter.

\section{Comparison to Dirac semimetals}
\label{cds}
\begin{figure} 
\includegraphics[width=8cm]{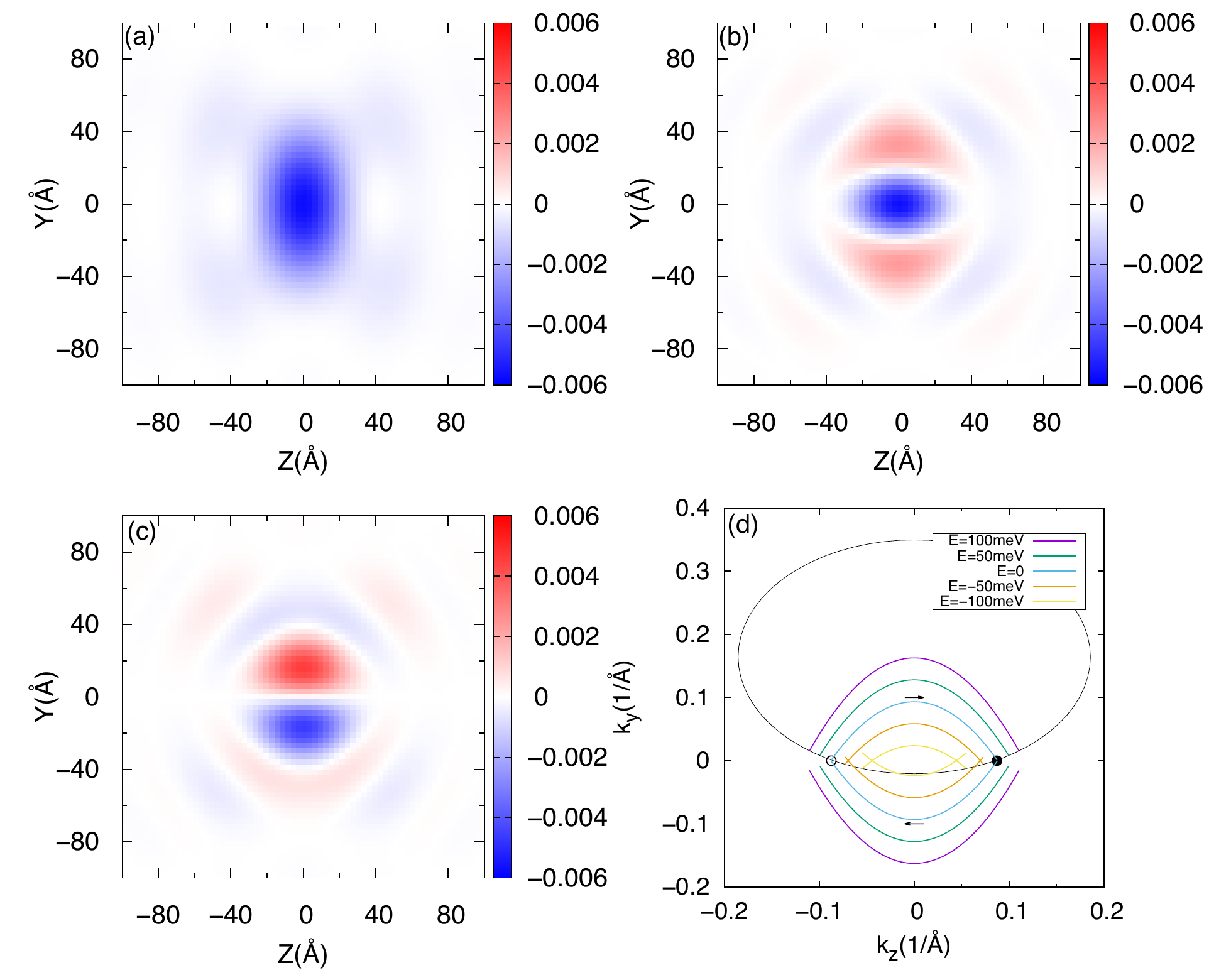}
\caption{Spatial spin-spin correlation functions (a) $J_{zz}$, (b) $J_{xx}$ and (c) $J_{xy}$ for (d) Dirac semimetal Fermi arcs in Na$_{3}$Bi, where the chemical potential is set at the Dirac points. Parameters are $\mu = 0 $eV, $\Delta = 0.3$eV, $\epsilon_{d} = -0.5$eV and $V_{\vec k}=0.8$eV. We take the $(100)$-plane and both two Dirac points (shown as circles) sit on the $k_{z}$ axis. The Fermi arcs consist of two sets of parabolas inside the ellipse. One set of parabola is spin-polarized along $k_{z}$ direction, the other along $-k_{z}$(shown by the blue arrows). The Fermi arcs of Dirac semimetal Na$_{3}$Bi have a higher symmetry so that $J_{xx}=J_{yy}$ and $J_{xz}=J_{zx}=J_{yz}=J_{zy}=0$. The shape and the spin texture of the Fermi arcs in Na$_{3}$Bi is simpler than that in the Weyl semimetal model. So are the spin-spin correlation functions, which have less structure. In (a), (b) and (c), $Y$ and $Z$ are spatial coordinates and are in the unit of $\AA$. In (d), $k_{z}$ and $k_{y}$ are in the unit of $\AA^{-1}$.}\label{fig:JNa3Bi}
\end{figure}
Comparing to Dirac semimetals which also host topological nontrivial surface Fermi arcs, the spin texture of Weyl semimetal Fermi arcs is more complex and results in the more anisotropic spin-spin correlation functions. As an example, we make use of the effective model for the Fermi arcs on $(100)$-surface in Dirac semimetal Na$_{3}$Bi\cite{PhysRevB.85.195320,PhysRevB.91.235138}. Following the same procedure, the Kondo screening cloud there is calculated. In our approximation, the small parameter $\alpha$ in the $k\cdot p$ model is omitted, so that the one sheet of Fermi arcs is spin-polarized in $+z$ direction while another polarized in $-z$. Both of the two Dirac points sit on the $k_{z}$ axis, see Fig.\ref{fig:JNa3Bi}(d). The Fermi arcs there consist of two branches of parabolas inside an ellipse.

The spin-spin correlation functions are shown in Fig.\ref{fig:JNa3Bi}(a)-(c), with much simpler pattern. $J_{zz}$ has a oval-shaped antiferromagnetic correlation core, while $J_{xx}$ appears antiferromagnetic near the impurity site with small ferromagnetic oscillation some distance away. Off-diagonal correlation functions are also simplified. Note that all off-diagonal correlation functions containing $z$ are zero, i.e., $J_{xz}=J_{zx}=J_{yz}=J_{zy}=0$, since we have taken a rough approximation in which the Fermi arcs are fully polarized in $z$ direction so that the $z$-spin component has no correlation with other spin components. In addition, $J_{yx}=-J_{xy}$. Higher symmetry is displayed, as $J_{xx}=J_{yy}$. The reason is that the spin is fully polarized in $z$ direction and it has no preference in $xy$-plane.

The comparison shows that the pattern and the symmetry of the correlation functions reflects different spin texture of Fermi arcs. As Fermi arcs in Dirac semimetals usually have simpler spin texture, this may help to distinguish Fermi arcs in Dirac semimetals from that in Weyl semimetals.

\section{Conclusion and Discussion}
\label{codi}
With the variational method, we have studied the Kondo effect of the Fermi arcs in a Weyl semimetal. The influence of the nontrivial spin texture is manifested in the spatial spin-spin correlation functions. We find that the correlation functions are highly anisotropic, in both real and spin space. The diagonal correlation functions feature regions of antiferromagnetic correlation near the impurity site while small regions of ferromagnetic correlation also exist. The complex pattern of off-diagonal correlation functions reflects the nontrivial spin texture. $J_{xx}$, $J_{yy}$, $J_{zz}$ and $J_{xy}$ all obey the $C_{2}$ rotational symmetry as the spin texture does, while the $J_{xz}$ and $J_{yz}$ get a minus sign when rotated by $\pi$ angle to respect time-reversal symmetry. When chemical potential is raised, the spin-spin correlation decays faster as more states in the Fermi arcs come to screen the impurity out. At the same time, the pattern of the correlation function changes according to the distribution of the filled Fermi arcs states. The screening cloud of the Weyl semimetal Fermi arcs and that of the Dirac semimetal Na$_{3}$Bi is distinguishable, since the latter has higher symmetry and simpler spin texture.

We have made some approximation throughout the calculation. The spin of the conducting electrons is taken as a good quantum number while actually it is not so at the presence of strong spin-orbit coupling. Further investigation may be conducted with an appropriate model. In our calculation, the contribution from the bulk states in the Weyl semimetals is omitted. On the one hand, we are considering the situation where the magnetic impurity is on the surface and it is expected that the interaction between the impurity and the surface states are much stronger. On the other hand, for the bulk states $\mu =0$ is a quantum critical point and a critical coupling strength exists below which the impurity remains unscreened. When the chemical potential is set at the Dirac/Weyl point, the Kondo screening comes only from the surface part.

\section{Acknowledgments}
We thank Qing-Feng Sun, Hua Jiang, Jin-hua Gao for helpful discussion. This work is financially supported by NBRPC (Grants No. 2015CB921102), NSFC (Grants No. 11534001, No. 11504008, and No. 11674028).

\bibliographystyle{apsrev4-1}
\bibliography{KFA}

\begin{thebibliography}{41}%
\makeatletter
\providecommand \@ifxundefined [1]{%
 \@ifx{#1\undefined}
}%
\providecommand \@ifnum [1]{%
 \ifnum #1\expandafter \@firstoftwo
 \else \expandafter \@secondoftwo
 \fi
}%
\providecommand \@ifx [1]{%
 \ifx #1\expandafter \@firstoftwo
 \else \expandafter \@secondoftwo
 \fi
}%
\providecommand \natexlab [1]{#1}%
\providecommand \enquote  [1]{``#1''}%
\providecommand \bibnamefont  [1]{#1}%
\providecommand \bibfnamefont [1]{#1}%
\providecommand \citenamefont [1]{#1}%
\providecommand \href@noop [0]{\@secondoftwo}%
\providecommand \href [0]{\begingroup \@sanitize@url \@href}%
\providecommand \@href[1]{\@@startlink{#1}\@@href}%
\providecommand \@@href[1]{\endgroup#1\@@endlink}%
\providecommand \@sanitize@url [0]{\catcode `\\12\catcode `\$12\catcode
  `\&12\catcode `\#12\catcode `\^12\catcode `\_12\catcode `\%12\relax}%
\providecommand \@@startlink[1]{}%
\providecommand \@@endlink[0]{}%
\providecommand \url  [0]{\begingroup\@sanitize@url \@url }%
\providecommand \@url [1]{\endgroup\@href {#1}{\urlprefix }}%
\providecommand \urlprefix  [0]{URL }%
\providecommand \Eprint [0]{\href }%
\providecommand \doibase [0]{http://dx.doi.org/}%
\providecommand \selectlanguage [0]{\@gobble}%
\providecommand \bibinfo  [0]{\@secondoftwo}%
\providecommand \bibfield  [0]{\@secondoftwo}%
\providecommand \translation [1]{[#1]}%
\providecommand \BibitemOpen [0]{}%
\providecommand \bibitemStop [0]{}%
\providecommand \bibitemNoStop [0]{.\EOS\space}%
\providecommand \EOS [0]{\spacefactor3000\relax}%
\providecommand \BibitemShut  [1]{\csname bibitem#1\endcsname}%
\let\auto@bib@innerbib\@empty
\bibitem [{\citenamefont {Weyl}(1929)}]{weyl1929elektron}%
  \BibitemOpen
  \bibfield  {author} {\bibinfo {author} {\bibfnamefont {H.}~\bibnamefont
  {Weyl}},\ }\href@noop {} {\bibfield  {journal} {\bibinfo  {journal} {Z.
  Phys.}\ }\textbf {\bibinfo {volume} {56}},\ \bibinfo {pages} {330} (\bibinfo
  {year} {1929})}\BibitemShut {NoStop}%
\bibitem [{\citenamefont {Wan}\ \emph {et~al.}(2011)\citenamefont {Wan},
  \citenamefont {Turner}, \citenamefont {Vishwanath},\ and\ \citenamefont
  {Savrasov}}]{PhysRevB.83.205101}%
  \BibitemOpen
  \bibfield  {author} {\bibinfo {author} {\bibfnamefont {X.}~\bibnamefont
  {Wan}}, \bibinfo {author} {\bibfnamefont {A.~M.}\ \bibnamefont {Turner}},
  \bibinfo {author} {\bibfnamefont {A.}~\bibnamefont {Vishwanath}}, \ and\
  \bibinfo {author} {\bibfnamefont {S.~Y.}\ \bibnamefont {Savrasov}},\ }\href
  {\doibase 10.1103/PhysRevB.83.205101} {\bibfield  {journal} {\bibinfo
  {journal} {Phys. Rev. B}\ }\textbf {\bibinfo {volume} {83}},\ \bibinfo
  {pages} {205101} (\bibinfo {year} {2011})}\BibitemShut {NoStop}%
\bibitem [{\citenamefont {Vafek}\ and\ \citenamefont
  {Vishwanath}(2014)}]{vafek2014dirac}%
  \BibitemOpen
  \bibfield  {author} {\bibinfo {author} {\bibfnamefont {O.}~\bibnamefont
  {Vafek}}\ and\ \bibinfo {author} {\bibfnamefont {A.}~\bibnamefont
  {Vishwanath}},\ }\href@noop {} {\bibfield  {journal} {\bibinfo  {journal}
  {Annu. Rev. Condens. Matter Phys.}\ }\textbf {\bibinfo {volume} {5}},\
  \bibinfo {pages} {83} (\bibinfo {year} {2014})}\BibitemShut {NoStop}%
\bibitem [{\citenamefont {Huang}\ \emph
  {et~al.}(2015{\natexlab{a}})\citenamefont {Huang}, \citenamefont {Xu},
  \citenamefont {Belopolski}, \citenamefont {Lee}, \citenamefont {Chang},
  \citenamefont {Wang}, \citenamefont {Alidoust}, \citenamefont {Bian},
  \citenamefont {Neupane}, \citenamefont {Zhang}, \citenamefont {Jia},
  \citenamefont {Bansil}, \citenamefont {Lin},\ and\ \citenamefont
  {Hasan}}]{Huang:2015aa}%
  \BibitemOpen
  \bibfield  {author} {\bibinfo {author} {\bibfnamefont {S.-M.}\ \bibnamefont
  {Huang}}, \bibinfo {author} {\bibfnamefont {S.-Y.}\ \bibnamefont {Xu}},
  \bibinfo {author} {\bibfnamefont {I.}~\bibnamefont {Belopolski}}, \bibinfo
  {author} {\bibfnamefont {C.-C.}\ \bibnamefont {Lee}}, \bibinfo {author}
  {\bibfnamefont {G.}~\bibnamefont {Chang}}, \bibinfo {author} {\bibfnamefont
  {B.}~\bibnamefont {Wang}}, \bibinfo {author} {\bibfnamefont {N.}~\bibnamefont
  {Alidoust}}, \bibinfo {author} {\bibfnamefont {G.}~\bibnamefont {Bian}},
  \bibinfo {author} {\bibfnamefont {M.}~\bibnamefont {Neupane}}, \bibinfo
  {author} {\bibfnamefont {C.}~\bibnamefont {Zhang}}, \bibinfo {author}
  {\bibfnamefont {S.}~\bibnamefont {Jia}}, \bibinfo {author} {\bibfnamefont
  {A.}~\bibnamefont {Bansil}}, \bibinfo {author} {\bibfnamefont
  {H.}~\bibnamefont {Lin}}, \ and\ \bibinfo {author} {\bibfnamefont {M.~Z.}\
  \bibnamefont {Hasan}},\ }\href {http://dx.doi.org/10.1038/ncomms8373}
  {\bibfield  {journal} {\bibinfo  {journal} {Nat. Commun.}\ }\textbf {\bibinfo
  {volume} {6}},\ \bibinfo {pages} {7373 EP } (\bibinfo {year}
  {2015}{\natexlab{a}})}\BibitemShut {NoStop}%
\bibitem [{\citenamefont {Lv}\ \emph {et~al.}(2015{\natexlab{a}})\citenamefont
  {Lv}, \citenamefont {Weng}, \citenamefont {Fu}, \citenamefont {Wang},
  \citenamefont {Miao}, \citenamefont {Ma}, \citenamefont {Richard},
  \citenamefont {Huang}, \citenamefont {Zhao}, \citenamefont {Chen},
  \citenamefont {Fang}, \citenamefont {Dai}, \citenamefont {Qian},\ and\
  \citenamefont {Ding}}]{PhysRevX.5.031013}%
  \BibitemOpen
  \bibfield  {author} {\bibinfo {author} {\bibfnamefont {B.~Q.}\ \bibnamefont
  {Lv}}, \bibinfo {author} {\bibfnamefont {H.~M.}\ \bibnamefont {Weng}},
  \bibinfo {author} {\bibfnamefont {B.~B.}\ \bibnamefont {Fu}}, \bibinfo
  {author} {\bibfnamefont {X.~P.}\ \bibnamefont {Wang}}, \bibinfo {author}
  {\bibfnamefont {H.}~\bibnamefont {Miao}}, \bibinfo {author} {\bibfnamefont
  {J.}~\bibnamefont {Ma}}, \bibinfo {author} {\bibfnamefont {P.}~\bibnamefont
  {Richard}}, \bibinfo {author} {\bibfnamefont {X.~C.}\ \bibnamefont {Huang}},
  \bibinfo {author} {\bibfnamefont {L.~X.}\ \bibnamefont {Zhao}}, \bibinfo
  {author} {\bibfnamefont {G.~F.}\ \bibnamefont {Chen}}, \bibinfo {author}
  {\bibfnamefont {Z.}~\bibnamefont {Fang}}, \bibinfo {author} {\bibfnamefont
  {X.}~\bibnamefont {Dai}}, \bibinfo {author} {\bibfnamefont {T.}~\bibnamefont
  {Qian}}, \ and\ \bibinfo {author} {\bibfnamefont {H.}~\bibnamefont {Ding}},\
  }\href {\doibase 10.1103/PhysRevX.5.031013} {\bibfield  {journal} {\bibinfo
  {journal} {Phys. Rev. X}\ }\textbf {\bibinfo {volume} {5}},\ \bibinfo {pages}
  {031013} (\bibinfo {year} {2015}{\natexlab{a}})}\BibitemShut {NoStop}%
\bibitem [{\citenamefont {Yang}\ \emph {et~al.}(2015)\citenamefont {Yang},
  \citenamefont {Liu}, \citenamefont {Sun}, \citenamefont {Peng}, \citenamefont
  {Yang}, \citenamefont {Zhang}, \citenamefont {Zhou}, \citenamefont {Zhang},
  \citenamefont {Guo}, \citenamefont {Rahn} \emph {et~al.}}]{yang2015weyl}%
  \BibitemOpen
  \bibfield  {author} {\bibinfo {author} {\bibfnamefont {L.}~\bibnamefont
  {Yang}}, \bibinfo {author} {\bibfnamefont {Z.}~\bibnamefont {Liu}}, \bibinfo
  {author} {\bibfnamefont {Y.}~\bibnamefont {Sun}}, \bibinfo {author}
  {\bibfnamefont {H.}~\bibnamefont {Peng}}, \bibinfo {author} {\bibfnamefont
  {H.}~\bibnamefont {Yang}}, \bibinfo {author} {\bibfnamefont {T.}~\bibnamefont
  {Zhang}}, \bibinfo {author} {\bibfnamefont {B.}~\bibnamefont {Zhou}},
  \bibinfo {author} {\bibfnamefont {Y.}~\bibnamefont {Zhang}}, \bibinfo
  {author} {\bibfnamefont {Y.}~\bibnamefont {Guo}}, \bibinfo {author}
  {\bibfnamefont {M.}~\bibnamefont {Rahn}},  \emph {et~al.},\ }\href@noop {}
  {\bibfield  {journal} {\bibinfo  {journal} {Nat. Phys.}\ }\textbf {\bibinfo
  {volume} {11}},\ \bibinfo {pages} {728} (\bibinfo {year} {2015})}\BibitemShut
  {NoStop}%
\bibitem [{\citenamefont {Xu}\ \emph {et~al.}(2015)\citenamefont {Xu},
  \citenamefont {Belopolski}, \citenamefont {Alidoust}, \citenamefont
  {Neupane}, \citenamefont {Bian}, \citenamefont {Zhang}, \citenamefont
  {Sankar}, \citenamefont {Chang}, \citenamefont {Yuan}, \citenamefont {Lee}
  \emph {et~al.}}]{xu2015discovery}%
  \BibitemOpen
  \bibfield  {author} {\bibinfo {author} {\bibfnamefont {S.-Y.}\ \bibnamefont
  {Xu}}, \bibinfo {author} {\bibfnamefont {I.}~\bibnamefont {Belopolski}},
  \bibinfo {author} {\bibfnamefont {N.}~\bibnamefont {Alidoust}}, \bibinfo
  {author} {\bibfnamefont {M.}~\bibnamefont {Neupane}}, \bibinfo {author}
  {\bibfnamefont {G.}~\bibnamefont {Bian}}, \bibinfo {author} {\bibfnamefont
  {C.}~\bibnamefont {Zhang}}, \bibinfo {author} {\bibfnamefont
  {R.}~\bibnamefont {Sankar}}, \bibinfo {author} {\bibfnamefont
  {G.}~\bibnamefont {Chang}}, \bibinfo {author} {\bibfnamefont
  {Z.}~\bibnamefont {Yuan}}, \bibinfo {author} {\bibfnamefont {C.-C.}\
  \bibnamefont {Lee}},  \emph {et~al.},\ }\href@noop {} {\bibfield  {journal}
  {\bibinfo  {journal} {Science}\ }\textbf {\bibinfo {volume} {349}},\ \bibinfo
  {pages} {613} (\bibinfo {year} {2015})}\BibitemShut {NoStop}%
\bibitem [{\citenamefont {Xu}\ \emph {et~al.}(2011)\citenamefont {Xu},
  \citenamefont {Weng}, \citenamefont {Wang}, \citenamefont {Dai},\ and\
  \citenamefont {Fang}}]{PhysRevLett.107.186806}%
  \BibitemOpen
  \bibfield  {author} {\bibinfo {author} {\bibfnamefont {G.}~\bibnamefont
  {Xu}}, \bibinfo {author} {\bibfnamefont {H.}~\bibnamefont {Weng}}, \bibinfo
  {author} {\bibfnamefont {Z.}~\bibnamefont {Wang}}, \bibinfo {author}
  {\bibfnamefont {X.}~\bibnamefont {Dai}}, \ and\ \bibinfo {author}
  {\bibfnamefont {Z.}~\bibnamefont {Fang}},\ }\href {\doibase
  10.1103/PhysRevLett.107.186806} {\bibfield  {journal} {\bibinfo  {journal}
  {Phys. Rev. Lett.}\ }\textbf {\bibinfo {volume} {107}},\ \bibinfo {pages}
  {186806} (\bibinfo {year} {2011})}\BibitemShut {NoStop}%
\bibitem [{\citenamefont {Burkov}\ and\ \citenamefont
  {Balents}(2011)}]{PhysRevLett.107.127205}%
  \BibitemOpen
  \bibfield  {author} {\bibinfo {author} {\bibfnamefont {A.~A.}\ \bibnamefont
  {Burkov}}\ and\ \bibinfo {author} {\bibfnamefont {L.}~\bibnamefont
  {Balents}},\ }\href {\doibase 10.1103/PhysRevLett.107.127205} {\bibfield
  {journal} {\bibinfo  {journal} {Phys. Rev. Lett.}\ }\textbf {\bibinfo
  {volume} {107}},\ \bibinfo {pages} {127205} (\bibinfo {year}
  {2011})}\BibitemShut {NoStop}%
\bibitem [{\citenamefont {Zyuzin}\ \emph {et~al.}(2012)\citenamefont {Zyuzin},
  \citenamefont {Wu},\ and\ \citenamefont {Burkov}}]{PhysRevB.85.165110}%
  \BibitemOpen
  \bibfield  {author} {\bibinfo {author} {\bibfnamefont {A.~A.}\ \bibnamefont
  {Zyuzin}}, \bibinfo {author} {\bibfnamefont {S.}~\bibnamefont {Wu}}, \ and\
  \bibinfo {author} {\bibfnamefont {A.~A.}\ \bibnamefont {Burkov}},\ }\href
  {\doibase 10.1103/PhysRevB.85.165110} {\bibfield  {journal} {\bibinfo
  {journal} {Phys. Rev. B}\ }\textbf {\bibinfo {volume} {85}},\ \bibinfo
  {pages} {165110} (\bibinfo {year} {2012})}\BibitemShut {NoStop}%
\bibitem [{\citenamefont {Hosur}\ and\ \citenamefont
  {Qi}(2013)}]{hosur2013recent}%
  \BibitemOpen
  \bibfield  {author} {\bibinfo {author} {\bibfnamefont {P.}~\bibnamefont
  {Hosur}}\ and\ \bibinfo {author} {\bibfnamefont {X.}~\bibnamefont {Qi}},\
  }\href@noop {} {\bibfield  {journal} {\bibinfo  {journal} {C. R. Physique}\
  }\textbf {\bibinfo {volume} {14}},\ \bibinfo {pages} {857} (\bibinfo {year}
  {2013})}\BibitemShut {NoStop}%
\bibitem [{\citenamefont {Burkov}(2014)}]{PhysRevLett.113.187202}%
  \BibitemOpen
  \bibfield  {author} {\bibinfo {author} {\bibfnamefont {A.~A.}\ \bibnamefont
  {Burkov}},\ }\href {\doibase 10.1103/PhysRevLett.113.187202} {\bibfield
  {journal} {\bibinfo  {journal} {Phys. Rev. Lett.}\ }\textbf {\bibinfo
  {volume} {113}},\ \bibinfo {pages} {187202} (\bibinfo {year}
  {2014})}\BibitemShut {NoStop}%
\bibitem [{\citenamefont {Wang}\ \emph {et~al.}(2016)\citenamefont {Wang},
  \citenamefont {Vergniory}, \citenamefont {Kushwaha}, \citenamefont
  {Hirschberger}, \citenamefont {Chulkov}, \citenamefont {Ernst}, \citenamefont
  {Ong}, \citenamefont {Cava},\ and\ \citenamefont
  {Bernevig}}]{PhysRevLett.117.236401}%
  \BibitemOpen
  \bibfield  {author} {\bibinfo {author} {\bibfnamefont {Z.}~\bibnamefont
  {Wang}}, \bibinfo {author} {\bibfnamefont {M.~G.}\ \bibnamefont {Vergniory}},
  \bibinfo {author} {\bibfnamefont {S.}~\bibnamefont {Kushwaha}}, \bibinfo
  {author} {\bibfnamefont {M.}~\bibnamefont {Hirschberger}}, \bibinfo {author}
  {\bibfnamefont {E.~V.}\ \bibnamefont {Chulkov}}, \bibinfo {author}
  {\bibfnamefont {A.}~\bibnamefont {Ernst}}, \bibinfo {author} {\bibfnamefont
  {N.~P.}\ \bibnamefont {Ong}}, \bibinfo {author} {\bibfnamefont {R.~J.}\
  \bibnamefont {Cava}}, \ and\ \bibinfo {author} {\bibfnamefont {B.~A.}\
  \bibnamefont {Bernevig}},\ }\href {\doibase 10.1103/PhysRevLett.117.236401}
  {\bibfield  {journal} {\bibinfo  {journal} {Phys. Rev. Lett.}\ }\textbf
  {\bibinfo {volume} {117}},\ \bibinfo {pages} {236401} (\bibinfo {year}
  {2016})}\BibitemShut {NoStop}%
\bibitem [{\citenamefont {Lv}\ \emph {et~al.}(2015{\natexlab{b}})\citenamefont
  {Lv}, \citenamefont {Xu}, \citenamefont {Weng}, \citenamefont {Ma},
  \citenamefont {Richard}, \citenamefont {Huang}, \citenamefont {Zhao},
  \citenamefont {Chen}, \citenamefont {Matt}, \citenamefont {Bisti} \emph
  {et~al.}}]{lv2015observation}%
  \BibitemOpen
  \bibfield  {author} {\bibinfo {author} {\bibfnamefont {B.}~\bibnamefont
  {Lv}}, \bibinfo {author} {\bibfnamefont {N.}~\bibnamefont {Xu}}, \bibinfo
  {author} {\bibfnamefont {H.}~\bibnamefont {Weng}}, \bibinfo {author}
  {\bibfnamefont {J.}~\bibnamefont {Ma}}, \bibinfo {author} {\bibfnamefont
  {P.}~\bibnamefont {Richard}}, \bibinfo {author} {\bibfnamefont
  {X.}~\bibnamefont {Huang}}, \bibinfo {author} {\bibfnamefont
  {L.}~\bibnamefont {Zhao}}, \bibinfo {author} {\bibfnamefont {G.}~\bibnamefont
  {Chen}}, \bibinfo {author} {\bibfnamefont {C.}~\bibnamefont {Matt}}, \bibinfo
  {author} {\bibfnamefont {F.}~\bibnamefont {Bisti}},  \emph {et~al.},\ }\href
  {\doibase 10.1038/nphys3426} {\bibfield  {journal} {\bibinfo  {journal} {Nat.
  Phys.}\ }\textbf {\bibinfo {volume} {11}},\ \bibinfo {pages} {724} (\bibinfo
  {year} {2015}{\natexlab{b}})}\BibitemShut {NoStop}%
\bibitem [{\citenamefont {Nielsen}\ and\ \citenamefont
  {Ninomiya}(1983)}]{nielsen1983adler}%
  \BibitemOpen
  \bibfield  {author} {\bibinfo {author} {\bibfnamefont {H.~B.}\ \bibnamefont
  {Nielsen}}\ and\ \bibinfo {author} {\bibfnamefont {M.}~\bibnamefont
  {Ninomiya}},\ }\href@noop {} {\bibfield  {journal} {\bibinfo  {journal}
  {Phys. Lett. B}\ }\textbf {\bibinfo {volume} {130}},\ \bibinfo {pages} {389}
  (\bibinfo {year} {1983})}\BibitemShut {NoStop}%
\bibitem [{\citenamefont {Liu}\ \emph {et~al.}(2013)\citenamefont {Liu},
  \citenamefont {Ye},\ and\ \citenamefont {Qi}}]{PhysRevB.87.235306}%
  \BibitemOpen
  \bibfield  {author} {\bibinfo {author} {\bibfnamefont {C.-X.}\ \bibnamefont
  {Liu}}, \bibinfo {author} {\bibfnamefont {P.}~\bibnamefont {Ye}}, \ and\
  \bibinfo {author} {\bibfnamefont {X.-L.}\ \bibnamefont {Qi}},\ }\href
  {\doibase 10.1103/PhysRevB.87.235306} {\bibfield  {journal} {\bibinfo
  {journal} {Phys. Rev. B}\ }\textbf {\bibinfo {volume} {87}},\ \bibinfo
  {pages} {235306} (\bibinfo {year} {2013})}\BibitemShut {NoStop}%
\bibitem [{\citenamefont {Son}\ and\ \citenamefont
  {Spivak}(2013)}]{PhysRevB.88.104412}%
  \BibitemOpen
  \bibfield  {author} {\bibinfo {author} {\bibfnamefont {D.~T.}\ \bibnamefont
  {Son}}\ and\ \bibinfo {author} {\bibfnamefont {B.~Z.}\ \bibnamefont
  {Spivak}},\ }\href {\doibase 10.1103/PhysRevB.88.104412} {\bibfield
  {journal} {\bibinfo  {journal} {Phys. Rev. B}\ }\textbf {\bibinfo {volume}
  {88}},\ \bibinfo {pages} {104412} (\bibinfo {year} {2013})}\BibitemShut
  {NoStop}%
\bibitem [{\citenamefont {Huang}\ \emph
  {et~al.}(2015{\natexlab{b}})\citenamefont {Huang}, \citenamefont {Zhao},
  \citenamefont {Long}, \citenamefont {Wang}, \citenamefont {Chen},
  \citenamefont {Yang}, \citenamefont {Liang}, \citenamefont {Xue},
  \citenamefont {Weng}, \citenamefont {Fang}, \citenamefont {Dai},\ and\
  \citenamefont {Chen}}]{PhysRevX.5.031023}%
  \BibitemOpen
  \bibfield  {author} {\bibinfo {author} {\bibfnamefont {X.}~\bibnamefont
  {Huang}}, \bibinfo {author} {\bibfnamefont {L.}~\bibnamefont {Zhao}},
  \bibinfo {author} {\bibfnamefont {Y.}~\bibnamefont {Long}}, \bibinfo {author}
  {\bibfnamefont {P.}~\bibnamefont {Wang}}, \bibinfo {author} {\bibfnamefont
  {D.}~\bibnamefont {Chen}}, \bibinfo {author} {\bibfnamefont {Z.}~\bibnamefont
  {Yang}}, \bibinfo {author} {\bibfnamefont {H.}~\bibnamefont {Liang}},
  \bibinfo {author} {\bibfnamefont {M.}~\bibnamefont {Xue}}, \bibinfo {author}
  {\bibfnamefont {H.}~\bibnamefont {Weng}}, \bibinfo {author} {\bibfnamefont
  {Z.}~\bibnamefont {Fang}}, \bibinfo {author} {\bibfnamefont {X.}~\bibnamefont
  {Dai}}, \ and\ \bibinfo {author} {\bibfnamefont {G.}~\bibnamefont {Chen}},\
  }\href {\doibase 10.1103/PhysRevX.5.031023} {\bibfield  {journal} {\bibinfo
  {journal} {Phys. Rev. X}\ }\textbf {\bibinfo {volume} {5}},\ \bibinfo {pages}
  {031023} (\bibinfo {year} {2015}{\natexlab{b}})}\BibitemShut {NoStop}%
\bibitem [{\citenamefont {Lu}\ and\ \citenamefont {Shen}(2016)}]{Lu2016}%
  \BibitemOpen
  \bibfield  {author} {\bibinfo {author} {\bibfnamefont {H.-Z.}\ \bibnamefont
  {Lu}}\ and\ \bibinfo {author} {\bibfnamefont {S.-Q.}\ \bibnamefont {Shen}},\
  }\href {\doibase 10.1007/s11467-016-0609-y} {\bibfield  {journal} {\bibinfo
  {journal} {Front. Phys.}\ }\textbf {\bibinfo {volume} {12}},\ \bibinfo
  {pages} {127201} (\bibinfo {year} {2016})}\BibitemShut {NoStop}%
\bibitem [{\citenamefont {Potter}\ \emph {et~al.}(2014)\citenamefont {Potter},
  \citenamefont {Kimchi},\ and\ \citenamefont
  {Vishwanath}}]{potter2014quantum}%
  \BibitemOpen
  \bibfield  {author} {\bibinfo {author} {\bibfnamefont {A.~C.}\ \bibnamefont
  {Potter}}, \bibinfo {author} {\bibfnamefont {I.}~\bibnamefont {Kimchi}}, \
  and\ \bibinfo {author} {\bibfnamefont {A.}~\bibnamefont {Vishwanath}},\
  }\href@noop {} {\bibfield  {journal} {\bibinfo  {journal} {Nat. Commun.}\
  }\textbf {\bibinfo {volume} {5}},\ \bibinfo {pages} {5161} (\bibinfo {year}
  {2014})}\BibitemShut {NoStop}%
\bibitem [{\citenamefont {Moll}\ \emph {et~al.}(2016)\citenamefont {Moll},
  \citenamefont {Nair}, \citenamefont {Helm}, \citenamefont {Potter},
  \citenamefont {Kimchi}, \citenamefont {Vishwanath},\ and\ \citenamefont
  {Analytis}}]{moll2016transport}%
  \BibitemOpen
  \bibfield  {author} {\bibinfo {author} {\bibfnamefont {P.~J.}\ \bibnamefont
  {Moll}}, \bibinfo {author} {\bibfnamefont {N.~L.}\ \bibnamefont {Nair}},
  \bibinfo {author} {\bibfnamefont {T.}~\bibnamefont {Helm}}, \bibinfo {author}
  {\bibfnamefont {A.~C.}\ \bibnamefont {Potter}}, \bibinfo {author}
  {\bibfnamefont {I.}~\bibnamefont {Kimchi}}, \bibinfo {author} {\bibfnamefont
  {A.}~\bibnamefont {Vishwanath}}, \ and\ \bibinfo {author} {\bibfnamefont
  {J.~G.}\ \bibnamefont {Analytis}},\ }\href {\doibase doi:10.1038/nature18276}
  {\bibfield  {journal} {\bibinfo  {journal} {Nature}\ }\textbf {\bibinfo
  {volume} {535}},\ \bibinfo {pages} {266} (\bibinfo {year}
  {2016})}\BibitemShut {NoStop}%
\bibitem [{\citenamefont {Jian}\ \emph {et~al.}(2015)\citenamefont {Jian},
  \citenamefont {Jiang},\ and\ \citenamefont {Yao}}]{PhysRevLett.114.237001}%
  \BibitemOpen
  \bibfield  {author} {\bibinfo {author} {\bibfnamefont {S.-K.}\ \bibnamefont
  {Jian}}, \bibinfo {author} {\bibfnamefont {Y.-F.}\ \bibnamefont {Jiang}}, \
  and\ \bibinfo {author} {\bibfnamefont {H.}~\bibnamefont {Yao}},\ }\href
  {\doibase 10.1103/PhysRevLett.114.237001} {\bibfield  {journal} {\bibinfo
  {journal} {Phys. Rev. Lett.}\ }\textbf {\bibinfo {volume} {114}},\ \bibinfo
  {pages} {237001} (\bibinfo {year} {2015})}\BibitemShut {NoStop}%
\bibitem [{\citenamefont {Jiang}\ \emph {et~al.}(2015)\citenamefont {Jiang},
  \citenamefont {Jiang}, \citenamefont {Liu}, \citenamefont {Sun},\ and\
  \citenamefont {Xie}}]{PhysRevLett.115.156602}%
  \BibitemOpen
  \bibfield  {author} {\bibinfo {author} {\bibfnamefont {Q.-D.}\ \bibnamefont
  {Jiang}}, \bibinfo {author} {\bibfnamefont {H.}~\bibnamefont {Jiang}},
  \bibinfo {author} {\bibfnamefont {H.}~\bibnamefont {Liu}}, \bibinfo {author}
  {\bibfnamefont {Q.-F.}\ \bibnamefont {Sun}}, \ and\ \bibinfo {author}
  {\bibfnamefont {X.~C.}\ \bibnamefont {Xie}},\ }\href {\doibase
  10.1103/PhysRevLett.115.156602} {\bibfield  {journal} {\bibinfo  {journal}
  {Phys. Rev. Lett.}\ }\textbf {\bibinfo {volume} {115}},\ \bibinfo {pages}
  {156602} (\bibinfo {year} {2015})}\BibitemShut {NoStop}%
\bibitem [{\citenamefont {Jiang}\ \emph {et~al.}(2016)\citenamefont {Jiang},
  \citenamefont {Jiang}, \citenamefont {Liu}, \citenamefont {Sun},\ and\
  \citenamefont {Xie}}]{PhysRevB.93.195165}%
  \BibitemOpen
  \bibfield  {author} {\bibinfo {author} {\bibfnamefont {Q.-D.}\ \bibnamefont
  {Jiang}}, \bibinfo {author} {\bibfnamefont {H.}~\bibnamefont {Jiang}},
  \bibinfo {author} {\bibfnamefont {H.}~\bibnamefont {Liu}}, \bibinfo {author}
  {\bibfnamefont {Q.-F.}\ \bibnamefont {Sun}}, \ and\ \bibinfo {author}
  {\bibfnamefont {X.~C.}\ \bibnamefont {Xie}},\ }\href {\doibase
  10.1103/PhysRevB.93.195165} {\bibfield  {journal} {\bibinfo  {journal} {Phys.
  Rev. B}\ }\textbf {\bibinfo {volume} {93}},\ \bibinfo {pages} {195165}
  (\bibinfo {year} {2016})}\BibitemShut {NoStop}%
\bibitem [{\citenamefont {Chen}\ \emph {et~al.}(2015)\citenamefont {Chen},
  \citenamefont {Song}, \citenamefont {Jiang}, \citenamefont {Sun},
  \citenamefont {Wang},\ and\ \citenamefont {Xie}}]{PhysRevLett.115.246603}%
  \BibitemOpen
  \bibfield  {author} {\bibinfo {author} {\bibfnamefont {C.-Z.}\ \bibnamefont
  {Chen}}, \bibinfo {author} {\bibfnamefont {J.}~\bibnamefont {Song}}, \bibinfo
  {author} {\bibfnamefont {H.}~\bibnamefont {Jiang}}, \bibinfo {author}
  {\bibfnamefont {Q.-F.}\ \bibnamefont {Sun}}, \bibinfo {author} {\bibfnamefont
  {Z.}~\bibnamefont {Wang}}, \ and\ \bibinfo {author} {\bibfnamefont {X.~C.}\
  \bibnamefont {Xie}},\ }\href {\doibase 10.1103/PhysRevLett.115.246603}
  {\bibfield  {journal} {\bibinfo  {journal} {Phys. Rev. Lett.}\ }\textbf
  {\bibinfo {volume} {115}},\ \bibinfo {pages} {246603} (\bibinfo {year}
  {2015})}\BibitemShut {NoStop}%
\bibitem [{\citenamefont {Liu}\ \emph {et~al.}(2016)\citenamefont {Liu},
  \citenamefont {Ohtsuki},\ and\ \citenamefont
  {Shindou}}]{PhysRevLett.116.066401}%
  \BibitemOpen
  \bibfield  {author} {\bibinfo {author} {\bibfnamefont {S.}~\bibnamefont
  {Liu}}, \bibinfo {author} {\bibfnamefont {T.}~\bibnamefont {Ohtsuki}}, \ and\
  \bibinfo {author} {\bibfnamefont {R.}~\bibnamefont {Shindou}},\ }\href
  {\doibase 10.1103/PhysRevLett.116.066401} {\bibfield  {journal} {\bibinfo
  {journal} {Phys. Rev. Lett.}\ }\textbf {\bibinfo {volume} {116}},\ \bibinfo
  {pages} {066401} (\bibinfo {year} {2016})}\BibitemShut {NoStop}%
\bibitem [{\citenamefont {Mitchell}\ and\ \citenamefont
  {Fritz}(2015)}]{PhysRevB.92.121109}%
  \BibitemOpen
  \bibfield  {author} {\bibinfo {author} {\bibfnamefont {A.~K.}\ \bibnamefont
  {Mitchell}}\ and\ \bibinfo {author} {\bibfnamefont {L.}~\bibnamefont
  {Fritz}},\ }\href {\doibase 10.1103/PhysRevB.92.121109} {\bibfield  {journal}
  {\bibinfo  {journal} {Phys. Rev. B}\ }\textbf {\bibinfo {volume} {92}},\
  \bibinfo {pages} {121109} (\bibinfo {year} {2015})}\BibitemShut {NoStop}%
\bibitem [{\citenamefont {Principi}\ \emph {et~al.}(2015)\citenamefont
  {Principi}, \citenamefont {Vignale},\ and\ \citenamefont
  {Rossi}}]{PhysRevB.92.041107}%
  \BibitemOpen
  \bibfield  {author} {\bibinfo {author} {\bibfnamefont {A.}~\bibnamefont
  {Principi}}, \bibinfo {author} {\bibfnamefont {G.}~\bibnamefont {Vignale}}, \
  and\ \bibinfo {author} {\bibfnamefont {E.}~\bibnamefont {Rossi}},\ }\href
  {\doibase 10.1103/PhysRevB.92.041107} {\bibfield  {journal} {\bibinfo
  {journal} {Phys. Rev. B}\ }\textbf {\bibinfo {volume} {92}},\ \bibinfo
  {pages} {041107} (\bibinfo {year} {2015})}\BibitemShut {NoStop}%
\bibitem [{\citenamefont {Sun}\ \emph {et~al.}(2015{\natexlab{a}})\citenamefont
  {Sun}, \citenamefont {Xu}, \citenamefont {Zhang},\ and\ \citenamefont
  {Zhou}}]{PhysRevB.92.195124}%
  \BibitemOpen
  \bibfield  {author} {\bibinfo {author} {\bibfnamefont {J.-H.}\ \bibnamefont
  {Sun}}, \bibinfo {author} {\bibfnamefont {D.-H.}\ \bibnamefont {Xu}},
  \bibinfo {author} {\bibfnamefont {F.-C.}\ \bibnamefont {Zhang}}, \ and\
  \bibinfo {author} {\bibfnamefont {Y.}~\bibnamefont {Zhou}},\ }\href {\doibase
  10.1103/PhysRevB.92.195124} {\bibfield  {journal} {\bibinfo  {journal} {Phys.
  Rev. B}\ }\textbf {\bibinfo {volume} {92}},\ \bibinfo {pages} {195124}
  (\bibinfo {year} {2015}{\natexlab{a}})}\BibitemShut {NoStop}%
\bibitem [{\citenamefont {Hosseini}\ and\ \citenamefont
  {Askari}(2015)}]{PhysRevB.92.224435}%
  \BibitemOpen
  \bibfield  {author} {\bibinfo {author} {\bibfnamefont {M.~V.}\ \bibnamefont
  {Hosseini}}\ and\ \bibinfo {author} {\bibfnamefont {M.}~\bibnamefont
  {Askari}},\ }\href {\doibase 10.1103/PhysRevB.92.224435} {\bibfield
  {journal} {\bibinfo  {journal} {Phys. Rev. B}\ }\textbf {\bibinfo {volume}
  {92}},\ \bibinfo {pages} {224435} (\bibinfo {year} {2015})}\BibitemShut
  {NoStop}%
\bibitem [{\citenamefont {Chang}\ \emph {et~al.}(2015)\citenamefont {Chang},
  \citenamefont {Zhou}, \citenamefont {Wang}, \citenamefont {Shan},\ and\
  \citenamefont {Xiao}}]{PhysRevB.92.241103}%
  \BibitemOpen
  \bibfield  {author} {\bibinfo {author} {\bibfnamefont {H.-R.}\ \bibnamefont
  {Chang}}, \bibinfo {author} {\bibfnamefont {J.}~\bibnamefont {Zhou}},
  \bibinfo {author} {\bibfnamefont {S.-X.}\ \bibnamefont {Wang}}, \bibinfo
  {author} {\bibfnamefont {W.-Y.}\ \bibnamefont {Shan}}, \ and\ \bibinfo
  {author} {\bibfnamefont {D.}~\bibnamefont {Xiao}},\ }\href {\doibase
  10.1103/PhysRevB.92.241103} {\bibfield  {journal} {\bibinfo  {journal} {Phys.
  Rev. B}\ }\textbf {\bibinfo {volume} {92}},\ \bibinfo {pages} {241103}
  (\bibinfo {year} {2015})}\BibitemShut {NoStop}%
\bibitem [{\citenamefont {Sun}\ \emph {et~al.}(2015{\natexlab{b}})\citenamefont
  {Sun}, \citenamefont {Wu},\ and\ \citenamefont {Yan}}]{PhysRevB.92.115428}%
  \BibitemOpen
  \bibfield  {author} {\bibinfo {author} {\bibfnamefont {Y.}~\bibnamefont
  {Sun}}, \bibinfo {author} {\bibfnamefont {S.-C.}\ \bibnamefont {Wu}}, \ and\
  \bibinfo {author} {\bibfnamefont {B.}~\bibnamefont {Yan}},\ }\href {\doibase
  10.1103/PhysRevB.92.115428} {\bibfield  {journal} {\bibinfo  {journal} {Phys.
  Rev. B}\ }\textbf {\bibinfo {volume} {92}},\ \bibinfo {pages} {115428}
  (\bibinfo {year} {2015}{\natexlab{b}})}\BibitemShut {NoStop}%
\bibitem [{\citenamefont {Lv}\ \emph {et~al.}(2015{\natexlab{c}})\citenamefont
  {Lv}, \citenamefont {Muff}, \citenamefont {Qian}, \citenamefont {Song},
  \citenamefont {Nie}, \citenamefont {Xu}, \citenamefont {Richard},
  \citenamefont {Matt}, \citenamefont {Plumb}, \citenamefont {Zhao},
  \citenamefont {Chen}, \citenamefont {Fang}, \citenamefont {Dai},
  \citenamefont {Dil}, \citenamefont {Mesot}, \citenamefont {Shi},
  \citenamefont {Weng},\ and\ \citenamefont {Ding}}]{PhysRevLett.115.217601}%
  \BibitemOpen
  \bibfield  {author} {\bibinfo {author} {\bibfnamefont {B.~Q.}\ \bibnamefont
  {Lv}}, \bibinfo {author} {\bibfnamefont {S.}~\bibnamefont {Muff}}, \bibinfo
  {author} {\bibfnamefont {T.}~\bibnamefont {Qian}}, \bibinfo {author}
  {\bibfnamefont {Z.~D.}\ \bibnamefont {Song}}, \bibinfo {author}
  {\bibfnamefont {S.~M.}\ \bibnamefont {Nie}}, \bibinfo {author} {\bibfnamefont
  {N.}~\bibnamefont {Xu}}, \bibinfo {author} {\bibfnamefont {P.}~\bibnamefont
  {Richard}}, \bibinfo {author} {\bibfnamefont {C.~E.}\ \bibnamefont {Matt}},
  \bibinfo {author} {\bibfnamefont {N.~C.}\ \bibnamefont {Plumb}}, \bibinfo
  {author} {\bibfnamefont {L.~X.}\ \bibnamefont {Zhao}}, \bibinfo {author}
  {\bibfnamefont {G.~F.}\ \bibnamefont {Chen}}, \bibinfo {author}
  {\bibfnamefont {Z.}~\bibnamefont {Fang}}, \bibinfo {author} {\bibfnamefont
  {X.}~\bibnamefont {Dai}}, \bibinfo {author} {\bibfnamefont {J.~H.}\
  \bibnamefont {Dil}}, \bibinfo {author} {\bibfnamefont {J.}~\bibnamefont
  {Mesot}}, \bibinfo {author} {\bibfnamefont {M.}~\bibnamefont {Shi}}, \bibinfo
  {author} {\bibfnamefont {H.~M.}\ \bibnamefont {Weng}}, \ and\ \bibinfo
  {author} {\bibfnamefont {H.}~\bibnamefont {Ding}},\ }\href {\doibase
  10.1103/PhysRevLett.115.217601} {\bibfield  {journal} {\bibinfo  {journal}
  {Phys. Rev. Lett.}\ }\textbf {\bibinfo {volume} {115}},\ \bibinfo {pages}
  {217601} (\bibinfo {year} {2015}{\natexlab{c}})}\BibitemShut {NoStop}%
\bibitem [{\citenamefont {Xu}\ \emph {et~al.}(2016)\citenamefont {Xu},
  \citenamefont {Belopolski}, \citenamefont {Sanchez}, \citenamefont {Neupane},
  \citenamefont {Chang}, \citenamefont {Yaji}, \citenamefont {Yuan},
  \citenamefont {Zhang}, \citenamefont {Kuroda}, \citenamefont {Bian},
  \citenamefont {Guo}, \citenamefont {Lu}, \citenamefont {Chang}, \citenamefont
  {Alidoust}, \citenamefont {Zheng}, \citenamefont {Lee}, \citenamefont
  {Huang}, \citenamefont {Hsu}, \citenamefont {Jeng}, \citenamefont {Bansil},
  \citenamefont {Neupert}, \citenamefont {Komori}, \citenamefont {Kondo},
  \citenamefont {Shin}, \citenamefont {Lin}, \citenamefont {Jia},\ and\
  \citenamefont {Hasan}}]{PhysRevLett.116.096801}%
  \BibitemOpen
  \bibfield  {author} {\bibinfo {author} {\bibfnamefont {S.-Y.}\ \bibnamefont
  {Xu}}, \bibinfo {author} {\bibfnamefont {I.}~\bibnamefont {Belopolski}},
  \bibinfo {author} {\bibfnamefont {D.~S.}\ \bibnamefont {Sanchez}}, \bibinfo
  {author} {\bibfnamefont {M.}~\bibnamefont {Neupane}}, \bibinfo {author}
  {\bibfnamefont {G.}~\bibnamefont {Chang}}, \bibinfo {author} {\bibfnamefont
  {K.}~\bibnamefont {Yaji}}, \bibinfo {author} {\bibfnamefont {Z.}~\bibnamefont
  {Yuan}}, \bibinfo {author} {\bibfnamefont {C.}~\bibnamefont {Zhang}},
  \bibinfo {author} {\bibfnamefont {K.}~\bibnamefont {Kuroda}}, \bibinfo
  {author} {\bibfnamefont {G.}~\bibnamefont {Bian}}, \bibinfo {author}
  {\bibfnamefont {C.}~\bibnamefont {Guo}}, \bibinfo {author} {\bibfnamefont
  {H.}~\bibnamefont {Lu}}, \bibinfo {author} {\bibfnamefont {T.-R.}\
  \bibnamefont {Chang}}, \bibinfo {author} {\bibfnamefont {N.}~\bibnamefont
  {Alidoust}}, \bibinfo {author} {\bibfnamefont {H.}~\bibnamefont {Zheng}},
  \bibinfo {author} {\bibfnamefont {C.-C.}\ \bibnamefont {Lee}}, \bibinfo
  {author} {\bibfnamefont {S.-M.}\ \bibnamefont {Huang}}, \bibinfo {author}
  {\bibfnamefont {C.-H.}\ \bibnamefont {Hsu}}, \bibinfo {author} {\bibfnamefont
  {H.-T.}\ \bibnamefont {Jeng}}, \bibinfo {author} {\bibfnamefont
  {A.}~\bibnamefont {Bansil}}, \bibinfo {author} {\bibfnamefont
  {T.}~\bibnamefont {Neupert}}, \bibinfo {author} {\bibfnamefont
  {F.}~\bibnamefont {Komori}}, \bibinfo {author} {\bibfnamefont
  {T.}~\bibnamefont {Kondo}}, \bibinfo {author} {\bibfnamefont
  {S.}~\bibnamefont {Shin}}, \bibinfo {author} {\bibfnamefont {H.}~\bibnamefont
  {Lin}}, \bibinfo {author} {\bibfnamefont {S.}~\bibnamefont {Jia}}, \ and\
  \bibinfo {author} {\bibfnamefont {M.~Z.}\ \bibnamefont {Hasan}},\ }\href
  {\doibase 10.1103/PhysRevLett.116.096801} {\bibfield  {journal} {\bibinfo
  {journal} {Phys. Rev. Lett.}\ }\textbf {\bibinfo {volume} {116}},\ \bibinfo
  {pages} {096801} (\bibinfo {year} {2016})}\BibitemShut {NoStop}%
\bibitem [{\citenamefont {Ojanen}(2013)}]{PhysRevB.87.245112}%
  \BibitemOpen
  \bibfield  {author} {\bibinfo {author} {\bibfnamefont {T.}~\bibnamefont
  {Ojanen}},\ }\href {\doibase 10.1103/PhysRevB.87.245112} {\bibfield
  {journal} {\bibinfo  {journal} {Phys. Rev. B}\ }\textbf {\bibinfo {volume}
  {87}},\ \bibinfo {pages} {245112} (\bibinfo {year} {2013})}\BibitemShut
  {NoStop}%
\bibitem [{\citenamefont {Yosida}(1966)}]{PhysRev.147.223}%
  \BibitemOpen
  \bibfield  {author} {\bibinfo {author} {\bibfnamefont {K.}~\bibnamefont
  {Yosida}},\ }\href {\doibase 10.1103/PhysRev.147.223} {\bibfield  {journal}
  {\bibinfo  {journal} {Phys. Rev.}\ }\textbf {\bibinfo {volume} {147}},\
  \bibinfo {pages} {223} (\bibinfo {year} {1966})}\BibitemShut {NoStop}%
\bibitem [{\citenamefont {Varma}\ and\ \citenamefont
  {Yafet}(1976)}]{PhysRevB.13.2950}%
  \BibitemOpen
  \bibfield  {author} {\bibinfo {author} {\bibfnamefont {C.~M.}\ \bibnamefont
  {Varma}}\ and\ \bibinfo {author} {\bibfnamefont {Y.}~\bibnamefont {Yafet}},\
  }\href {\doibase 10.1103/PhysRevB.13.2950} {\bibfield  {journal} {\bibinfo
  {journal} {Phys. Rev. B}\ }\textbf {\bibinfo {volume} {13}},\ \bibinfo
  {pages} {2950} (\bibinfo {year} {1976})}\BibitemShut {NoStop}%
\bibitem [{\citenamefont {Feng}\ \emph {et~al.}(2010)\citenamefont {Feng},
  \citenamefont {Chen}, \citenamefont {Gao}, \citenamefont {Wang},\ and\
  \citenamefont {Zhang}}]{PhysRevB.81.235411}%
  \BibitemOpen
  \bibfield  {author} {\bibinfo {author} {\bibfnamefont {X.-Y.}\ \bibnamefont
  {Feng}}, \bibinfo {author} {\bibfnamefont {W.-Q.}\ \bibnamefont {Chen}},
  \bibinfo {author} {\bibfnamefont {J.-H.}\ \bibnamefont {Gao}}, \bibinfo
  {author} {\bibfnamefont {Q.-H.}\ \bibnamefont {Wang}}, \ and\ \bibinfo
  {author} {\bibfnamefont {F.-C.}\ \bibnamefont {Zhang}},\ }\href {\doibase
  10.1103/PhysRevB.81.235411} {\bibfield  {journal} {\bibinfo  {journal} {Phys.
  Rev. B}\ }\textbf {\bibinfo {volume} {81}},\ \bibinfo {pages} {235411}
  (\bibinfo {year} {2010})}\BibitemShut {NoStop}%
\bibitem [{\citenamefont {Feng}\ and\ \citenamefont
  {Zhang}(2011)}]{feng2011kondo}%
  \BibitemOpen
  \bibfield  {author} {\bibinfo {author} {\bibfnamefont {X.-Y.}\ \bibnamefont
  {Feng}}\ and\ \bibinfo {author} {\bibfnamefont {F.-C.}\ \bibnamefont
  {Zhang}},\ }\href@noop {} {\bibfield  {journal} {\bibinfo  {journal} {J.
  Phys.: Condens. Matter}\ }\textbf {\bibinfo {volume} {23}},\ \bibinfo {pages}
  {105602} (\bibinfo {year} {2011})}\BibitemShut {NoStop}%
\bibitem [{\citenamefont {Wang}\ \emph {et~al.}(2012)\citenamefont {Wang},
  \citenamefont {Sun}, \citenamefont {Chen}, \citenamefont {Franchini},
  \citenamefont {Xu}, \citenamefont {Weng}, \citenamefont {Dai},\ and\
  \citenamefont {Fang}}]{PhysRevB.85.195320}%
  \BibitemOpen
  \bibfield  {author} {\bibinfo {author} {\bibfnamefont {Z.}~\bibnamefont
  {Wang}}, \bibinfo {author} {\bibfnamefont {Y.}~\bibnamefont {Sun}}, \bibinfo
  {author} {\bibfnamefont {X.-Q.}\ \bibnamefont {Chen}}, \bibinfo {author}
  {\bibfnamefont {C.}~\bibnamefont {Franchini}}, \bibinfo {author}
  {\bibfnamefont {G.}~\bibnamefont {Xu}}, \bibinfo {author} {\bibfnamefont
  {H.}~\bibnamefont {Weng}}, \bibinfo {author} {\bibfnamefont {X.}~\bibnamefont
  {Dai}}, \ and\ \bibinfo {author} {\bibfnamefont {Z.}~\bibnamefont {Fang}},\
  }\href {\doibase 10.1103/PhysRevB.85.195320} {\bibfield  {journal} {\bibinfo
  {journal} {Phys. Rev. B}\ }\textbf {\bibinfo {volume} {85}},\ \bibinfo
  {pages} {195320} (\bibinfo {year} {2012})}\BibitemShut {NoStop}%
\bibitem [{\citenamefont {Gorbar}\ \emph {et~al.}(2015)\citenamefont {Gorbar},
  \citenamefont {Miransky}, \citenamefont {Shovkovy},\ and\ \citenamefont
  {Sukhachov}}]{PhysRevB.91.235138}%
  \BibitemOpen
  \bibfield  {author} {\bibinfo {author} {\bibfnamefont {E.~V.}\ \bibnamefont
  {Gorbar}}, \bibinfo {author} {\bibfnamefont {V.~A.}\ \bibnamefont
  {Miransky}}, \bibinfo {author} {\bibfnamefont {I.~A.}\ \bibnamefont
  {Shovkovy}}, \ and\ \bibinfo {author} {\bibfnamefont {P.~O.}\ \bibnamefont
  {Sukhachov}},\ }\href {\doibase 10.1103/PhysRevB.91.235138} {\bibfield
  {journal} {\bibinfo  {journal} {Phys. Rev. B}\ }\textbf {\bibinfo {volume}
  {91}},\ \bibinfo {pages} {235138} (\bibinfo {year} {2015})}\BibitemShut
  {NoStop}%
\end{thebibliography}%

\end{document}